\documentclass[aps,pra,twocolumn,superscriptaddress,showpacs]{revtex4}
\usepackage{amsmath}
\usepackage{amssymb}
\usepackage{graphicx}

\begin{document}

\title{Study of the induced potential produced by ultrashort pulses on metal
surfaces }
\author{M. N. Faraggi}
\affiliation{Donostia International Physics Center (DIPC) Manuel de Lardizabal 4, San
Sebasti\'{a}n, Spain.}
\author{I. Aldazabal}
\affiliation{Donostia International Physics Center (DIPC) Manuel de Lardizabal 4, San
Sebasti\'{a}n, Spain.}
\affiliation{Centro de F\'{\i}sica de Materiales CSIC-UPV/EHU, San Sebasti\'{a}n, Spain.}
\author{M. S. Gravielle }
\affiliation{Instituto de Astronom\'{\i}a y F\'{\i}sica del Espacio, CONICET, C. C. 67,
Suc. 28, 1428 Buenos Aires, Argentina. Depto de F\'{\i}sica, FCEN-UBA,
Argentina.}
\author{A. Arnau}
\affiliation{Departamento de Fisica de Materiales, Facultad de Quimica UPV/EHU, San
Sebastian, Spain.}
\affiliation{Centro de F\'{\i}sica de Materiales CSIC-UPV/EHU, San Sebasti\'{a}n, Spain.}
\author{V. M. Silkin}
\affiliation{Donostia International Physics Center (DIPC) Manuel de Lardizabal 4, San
Sebasti\'{a}n, Spain.}
\affiliation{Centro de F\'{\i}sica de Materiales CSIC-UPV/EHU, San Sebasti\'{a}n, Spain.}
\date{\today }

\begin{abstract}
The influence of the induced potential on photoelectron emission from metal
surfaces is studied for grazing incidence of ultra short laser pulses. To
describe this process we introduce a distorted wave-method, the Surface
Jellium-Volkov approach, which includes the perturbation on the emitted
electron produced by both the laser and the induced fields. The method is
applied to an Al(111) surface contrasting the results with the numerical
solution to the time-dependent Schr\"{o}dinger equation (TDSE). We found
that SJV\ approach reproduces well the main features of emission spectra,
accounting properly for effects originated by the induced potential.
\end{abstract}

\maketitle

\section{Introduction}

In the past few years developments in laser technology have made it possible
to produce laser pulses with durations in the sub-femtosecond scale \cite%
{cava07,Gou07,Kien04,Hawo07, Baltu03}. This advance in the experimental area
opens up new branches in the research of the matter-radiation system \cite%
{Lorin08,Milo06,Kling08,Miaja08, Leme09}. In particular, the investigation
of photoelectron emission from surfaces due to incidence of short laser
pulses gives the chance to understand\ a piece of the complicated puzzle
corresponding to electron dynamics at metal surfaces.

In this article we investigate the photoelectron emission produced when an
ultrashort laser pulse impinges grazingly on a metal surface, focusing the
attention on the role played by the surface induced potential. \ The induced
potential is caused by the rearrangement of valence-band electrons due to
the presence of the external electromagnetic field. This potential is
expected not to affect appreciably electron emission for high frequencies of
the laser pulse, for which surface electrons are not able to follow the fast
fluctuations of the field. But for frequencies of the pulse close or lower
than the surface plasmon frequency, the induced potential becomes comparable
to the laser perturbation and its effect cannot be neglected. With this goal
we introduce a simple model, named \textit{Surface Jellium-Volkov} (SJV)
approximation, which includes information about the action of the surface
induced potential, taking into account the main features of the process.

The SJV approach is a time-dependent distorted wave method that makes use of
the well-known Volkov phase \cite{Volk35} to describe the interaction of the
active electron with the laser and the induced fields, while the surface
potential is represented within the jellium model. This kind of one-active
electron theories has been recently applied to study different laser-induced
electron emission processes from metal surfaces, providing reasonable
predictions \cite{Fais05,Fara07,Bagge08,Zhang09}. To corroborate the
validity of the proposed approximation, we compare SJV results with the
numerical solution of the time-dependent Schr\"{o}dinger equation (TDSE), in
which the contribution of the surface induced potential is also included.
The induced potential is here obtained from a linear response theory by
considering a jellium model for a one-dimensional slab.

With both methods - SJV and TDSE - we calculate the probability of electron
emission from the valence band of an Al surface, considering different
frequencies and durations of the laser pulse. We analyze in detail the
effect of the induced potential on electron distributions by comparing to
values derived from the previous \textit{Impulsive Jellium-Volkov} (IJV)
approximation \cite{Fara07}, which does not contain the dynamic response of
the surface.

The article is organized as follows. In Sec. II we present the theory, in
Sec. III results are shown and discussed, and finally in Sec. IV conclusions
are summarized. Atomic units are used throughout unless otherwise stated. \
\ \ \ 

\section{Theory}

Let us consider a laser pulse impinging grazingly on a metal surface $(S).$
As a consequence of the interaction, an electron $(e)$ of the valence band
of the solid, initially in the state $i$, is ejected above the vacuum level,
ending in a final state $f$. \ The frame of reference is placed at the
position of the crystal border, with the $\hat{z}$ axis perpendicular to the
surface, pointing towards the vacuum region.

For this collision system we can write the Hamiltonian corresponding to the
interacting electron as:%
\begin{equation}
H=H_{0}+V_{L}+V_{ind},  \label{Hamiltonian}
\end{equation}%
where $H_{0}=-\nabla _{\mathbf{r}}^{2}/2+V_{S}$ is the unperturbed
Hamiltonian, with $V_{S}$ the electron-surface potential, and $V_{L}=\mathbf{%
r.F}(t)$ represents the electron interaction with the laser field $\mathbf{F}%
(t)$ at the time $t$, expressed in the length gauge. In Eq.(\ref{Hamiltonian}%
), $V_{ind}$ denotes the surface induced potential, which is originated by
electronic density fluctuations produced by the external field.

The electron interaction with the surface, $V_{S}$, is here described with
the jellium model, being $V_{S}=-V_{c}\Theta (-z)$ with $V_{c}=E_{F}+E_{W}$,
where $E_{F}$ is the Fermi energy, $E_{W}$ is the work function and $\Theta $
denotes the unitary Heaviside function. This simple surface model has proved
to give an adequate description of the electron-surface interaction for
electron excitations from\ the valence band of metal surfaces \cite%
{Grav02,Thumm97, Fara07, Bagge08,Zhang09}. Within the jellium model the\
unperturbed electronic states, eigenstates of $H_{0}$, are written as: 
\begin{equation}
\Phi _{\mathbf{k}}^{\pm }(\mathbf{r},t)=\frac{e^{i\mathbf{k}_{s}.\mathbf{r}%
_{s}}}{2\pi }\phi _{k_{z}}^{\pm }(z)e^{-iE_{\mathbf{k}}t},  \label{eigfH0}
\end{equation}%
where the position vector of the active electron $e$ is expressed as $%
\mathbf{r\equiv (r}_{s},z\mathbf{)}$, with $\mathbf{r}_{s}$ and $z$ the
components parallel and perpendicular to the surface, respectively. The
vector $\mathbf{k=(k}_{s},k_{z}\mathbf{)}$ is the momentum measured inside
the solid and $E_{\mathbf{k}}$ $=k_{s}^{2}/2+\epsilon _{k_{z}}$ corresponds
to the electron energy. The signs $\pm $ define the outgoing (+) and
incoming (-) asymptotic conditions of the collision problem and the
eigenfunctions $\phi _{k_{z}}^{\pm }(z)$ with eigenenergy $\epsilon _{k_{z}}$
are given in the Appendix of Ref. \cite{Grav98}.

Taking into account the grazing incidence condition, together with the
translational invariance of $V_{S}$\ in the direction parallel to the
surface, we choose the electric field $\mathbf{F}\left( t\right) $
perpendicular to the surface plane, that is, along the $\hat{z}$-axis. The
temporal profile of the pulse is defined as:

\begin{equation}
F\left( t\right) =F_{0}\sin (\omega t+\varphi )\sin ^{2}(\pi t/\tau )\;\;
\label{F0}
\end{equation}%
for $0<t<\tau $ and $0$ elsewhere, where $F_{0}$ is the maximum field
strength, $\omega $ is the carrier frequency, $\varphi $ represents the
carrier envelope phase, and $\tau $ determines the duration of the pulse.

The differential probability of electron emission from the surface is
expressed in terms of the transition matrix as:

\begin{equation}
\frac{dP}{d\mathbf{k}_{f}^{\prime }}=\rho _{e}\frac{k_{fz}^{\prime }}{k_{fz}}%
\int d\mathbf{k}_{i}\ \Theta (v_{F}-k_{i})\ \left\vert T_{if}\right\vert
^{2},  \label{dP}
\end{equation}%
where $T_{if}$ is T-matrix element corresponding to the inelastic transition 
$\mathbf{k}_{i}\mathbf{\rightarrow k}_{f}^{\prime }$ and $\mathbf{k}%
_{f}^{\prime }=\mathbf{(k}_{fs},k_{fz}^{\prime }\mathbf{)}$ is the final
electron momentum outside the solid, with $\ k_{fz}^{\prime }=\left( \
k_{fz}^{2}\ -2V_{c}\right) ^{1/2}$. In Eq. (\ref{dP}), $\rho _{e}=2$ takes
into account the spin states and $\Theta $ restricts the initial states to
those contained within the Fermi sphere, with $v_{F}=(2E_{F})^{1/2}$. The
angular distribution of emitted electron can be derived in a straightforward
way from Eq. (\ref{dP}) as $d^{2}P/dE_{f}d\Omega _{f}=k_{f}^{\prime }\ dP/d%
\mathbf{k}_{f}^{\prime },$ where $E_{f}$ and $\Omega _{f}$ are the final
energy and solid angle, respectively, of the ejected electron and $%
k_{f}^{\prime }=\left\vert \mathbf{k}_{f}^{\prime }\right\vert $.

In this work we evaluate $T_{if}$ by using two different methods: the SJV
approximation and the numerical solution of the TDSE. Both of them are
summarized below.

\subsection{Surface Jellium-Volkov approximation}

\ In the SJV theory, the final distorted state is represented by the Surface
Jellium Volkov wave function, which includes the actions of the laser field
and the induced potential on the emitted electron, both described by means
of the Volkov phase. The induced potential is derived from a linear response
theory by using a one-dimensional jellium model \cite{Aldu03}. It can be
expressed as $V_{ind}(z\mathbf{,}t)=z$ $g(t)$ inside the solid, with the
function $g(t)$ numerically determined, while outside the solid - in the
vacuum region- $V_{ind}(z,t)=0$. Hence, the final SJV wave function can be
written as%
\begin{equation}
\chi _{f}^{(SJV)-}(\mathbf{r},t)=\Phi _{\mathbf{k}_{f}}^{-}(\mathbf{r}%
,t)\exp \left[ iD_{L}^{-}(k_{fz},z,t)\right] \xi _{ind}(z,t),  \label{fi3bis}
\end{equation}%
where $\Phi _{\mathbf{k}_{f}}^{-}(\mathbf{r},t)$ is the unperturbed final
state given by Eq.(\ref{eigfH0}), which includes the asymptotic condition
corresponding to emission towards the vacuum zone (external ionization
process \cite{Grav98}). In Eq. (\ref{fi3bis}), the function $D_{L}^{-}$
represents the Volkov phase associated with the laser field, which is
expressed as:

\begin{equation}
D_{L}^{-}(k_{fz},z,t)=\frac{z}{c}\ A^{-}(t)-\beta ^{-}(t)-k_{fz}\ \alpha
^{-}(t).  \label{phase}
\end{equation}%
The temporal functions involved in Eq. (\ref{phase}) are related to the
vector potential $A^{-}(t)$, the ponderomotive energy $\beta ^{-}(t)$ and
the quiver amplitude $\alpha ^{-}(t)$ of the pulse, being defined as:%
\begin{eqnarray}
A^{-}(t) &=&-c\int\nolimits_{+\infty }^{t}dt^{\prime }F(t^{\prime }), 
\notag \\
\beta ^{-}(t) &=&(2c^{2})^{-1}\int\nolimits_{+\infty }^{t}dt^{\prime
}[A^{-}(t^{\prime })]^{2},  \label{A1} \\
\alpha ^{-}(t) &=&c^{-1}\int\nolimits_{+\infty }^{t}dt^{\prime
}A^{-}(t^{\prime }),  \notag
\end{eqnarray}%
with $c$ the speed of light. In a similar way we express the function $\xi
_{ind}$, which considers the action of the induced potential on the active
electron, as

\begin{equation}
\xi _{ind}(z,t)=\left\{ 
\begin{array}{c}
\exp [i(z/c)A_{ind}^{-}(t)]\ \ \ \ \ \ \ \ \ \ \mathrm{for}\text{ \ \ \ }%
z\leq 0 \\ 
1\ \ \ \ \ \ \ \ \ \ \ \ \ \ \ \ \ \ \ \ \ \ \ \ \ \ \ \ \ \ \ \ \ \ \mathrm{%
for}\text{ \ \ \ }z>0%
\end{array}%
\right. ,  \label{find}
\end{equation}%
with $A_{ind}^{-}=-c\int\limits_{+\infty }^{t}dt^{\prime }\ g(t^{\prime })$
the momentum transferred by the induced field. Note that the effect of the
image charge of the emitted electron was not taken into account in the final
distorted wave function $\chi _{f}^{(SJV)-}$ because its contribution has
been found negligible \cite{Dombi06}.

Employing the final SJV wave function given by Eq. (\ref{fi3bis}) within a
time-dependent distorted-wave formalism \cite{Dewa94}, the transition
amplitude reads:

\begin{equation}
T_{if}^{(SJV)}=T^{(C)}+T^{(PC)},  \label{Tcijv}
\end{equation}%
where 
\begin{equation}
T^{(C)}=-i\int\limits_{0}^{\tau }dt\left\langle \chi
_{f}^{(SJV)-}(t)\left\vert U(t)\right\vert \Phi _{\mathbf{k}%
_{i}}^{+}(t)\right\rangle  \label{Tp}
\end{equation}%
represents the primary or collision (C) term, with $%
U(z,t)=V_{L}(z,t)+V_{ind}(z,t)$ the perturbation introduced by the laser and
the induced fields and $\Phi _{\mathbf{k}_{i}}^{+}$ the unperturbed initial
state, given by Eq. (\ref{eigfH0}). The second term of Eq. (\ref{Tcijv}), $%
T^{(PC)}$, is here called post-collision (PC) transition amplitude,
corresponding to the emission process after the pulse turns off at the time $%
\tau $. It reads:%
\begin{equation}
T^{(PC)}=-i\int\limits_{\tau }^{+\infty }dt\left\langle \chi
_{f}^{(SJV)-}(t)\left\vert V_{ind}(t)\right\vert \Phi _{\mathbf{k}%
_{i}}^{+}(t)\right\rangle .  \label{Tind1}
\end{equation}

\subsection{TDSE solution}

Replacing the semi-infinite jellium potential by the one corresponding to a
one-dimensional slab of size $a$, $V_{slab}=-V_{c}\ \Theta (a/2-z)\ \Theta
(a/2+z)$, and taking into account the symmetry of the system in the
direction parallel to the surface, we can write the unperturbed eigenstates
as 
\begin{equation}
\Phi _{\mathbf{k},n}(\mathbf{r},t)=\frac{e^{i\mathbf{k}_{s}.\mathbf{r}_{s}}}{%
2\pi }\varphi _{n}(z)e^{-iE_{\mathbf{k}}t},  \label{eq:slab_eigen}
\end{equation}%
where now the functions $\varphi _{n}(z)$ are the discretized
one-dimensional eigenstates of the slab potential.

The time evolution of the electronic eigenstates under the laser pulse
perturbation is governed by the one-dimensional time-dependent Schr\"{o}%
dinger equation 
\begin{equation}
i\frac{\partial }{\partial t}\varphi _{n}(z,t)=H(z,t)\varphi _{n}(z,t),
\end{equation}%
where the unperturbed part of the Hamiltonian $H(z,t)$ is now $%
H_{0}=-(1/2)(d^{2}/dz^{2})+V_{slab}$.

The discrete time step evolution is given by the evolution operator 
\begin{equation}
\varphi _{n}(z,t+\Delta t)=exp(-i\Delta tH)\varphi _{n}(z,t),
\end{equation}%
which is computed by using the Crank-Nicholson scheme, approximating the
exponential by the Cayley form \cite{Press92} 
\begin{equation}
exp(-i\Delta tH)\approx \frac{1-\frac{i\Delta t}{2}H}{1+\frac{i\Delta t}{2}H}%
.
\end{equation}%
This scheme is unitary, unconditionally stable, and accurate up to order $%
(H\Delta t)^{2}$.

To obtain the transition amplitude we evolve every eigenstate within the
Fermi sphere of the unperturbed slab, projecting then the evolved states
over the discretization box \textquotedblleft continuum\textquotedblright\
states, $\varphi _{f}^{k}(z)$, 
\begin{equation}
T_{if}^{(TDSE)}=\left\langle \varphi _{f}^{k}(z)|\varphi _{i}(z,t\rightarrow
\infty )\right\rangle .
\end{equation}%
Independence of the results with different slab sizes guarantees that the
used slab size accurately represents the semi-infinite medium. For the
simulation box we have taken completely reflective walls as boundary
conditions.

\bigskip

\section{Results}

We applied the SJV and TDSE\ methods to study electron emission from the
valence band of an Al(111) surface produced as a consequence of grazing
incidence of ultrashort and intense laser pulses. As Aluminum is a typical
metal surface, it will be considered as a benchmark for the theory. The
Al(111) is described by the following parameters: the Fermi energy $%
E_{F}=0.414$ $\mathrm{a.u.}$, the work function $E_{W}=0.156$ \textrm{a.u.},
and the surface plasmon frequency $\omega _{s}=0.4$ \textrm{a.u..}

For the TDSE calculations a slab with a width of $311.54$ \textrm{a.u.} ($%
142 $ Aluminum atomic layers), surrounded by $244.23$ \textrm{a.u.} of
vacuum on each side, was used. The grid sizes were $\Delta z=0.1$ a.u. for
the spacial grid and $\Delta t=0.005$ a.u. for the time grid. The time
evolution was considered finished when the induced potential had decayed two
orders of magnitude from its value at the moment the laser pulse was
switched off, at $t=\tau $. The same criteria was used to evaluate the upper
limit of the time integral of Eq. (\ref{Tind1}). Note that to compare the
SJV and TDSE results it is necessary to take into account that the former
theory includes the proper asymptotic conditions, distinguishing the
external from the internal ionization processes, while the latter does not.
Then, as a first estimation we weighted TDSE values with the fraction of
electrons emitted towards the vacuum derived from the SJV model \cite{Fara07}%
.

In this work we considered symmetric pulses, with $\varphi =-\omega \tau
/2+\pi /2$ . The field strength was fixed as $F_{0}=$ $0.001\ \mathrm{a.u.(}%
I\simeq $ $4$ $10^{10}$ \textrm{W/cm}$^{2}$\textrm{)}, which belongs to the
perturbative regime, far from the saturation region and the damage threshold 
\cite{Miaja08b, Miaja09}.\ In accord with results of a previous theory \cite%
{Fara07}, the maximum of the emission probability corresponds to the angle $%
\theta _{e}=90%
{{}^o}%
$, which coincides with the orientation of the laser field. Therefore, all
results presented here refer to this emission angle.

Since the dynamic response of the surface is characterized by the surface
plasmon frequency \ $\omega _{s}$, in order to investigate the influence of
\ the induced potential we varied the carrier frequency $\omega $ of the
laser field around the value of \ $\omega _{s}$. We start considering laser
pulses with several oscillations inside the envelope function, which
correspond to the so-called multiphoton regime. In this regime, related to a
Keldysh parameter $\gamma =\omega \sqrt{E_{W}}/F_{0}$ \cite{Keld65} greater
than the unity, the laser frequency tends to the photon energy and the
electron spectrum displays maxima associated with the absorption of photons.

\begin{figure}[!htb]
         \begin{center}
                     \includegraphics*[width=.45\textwidth]{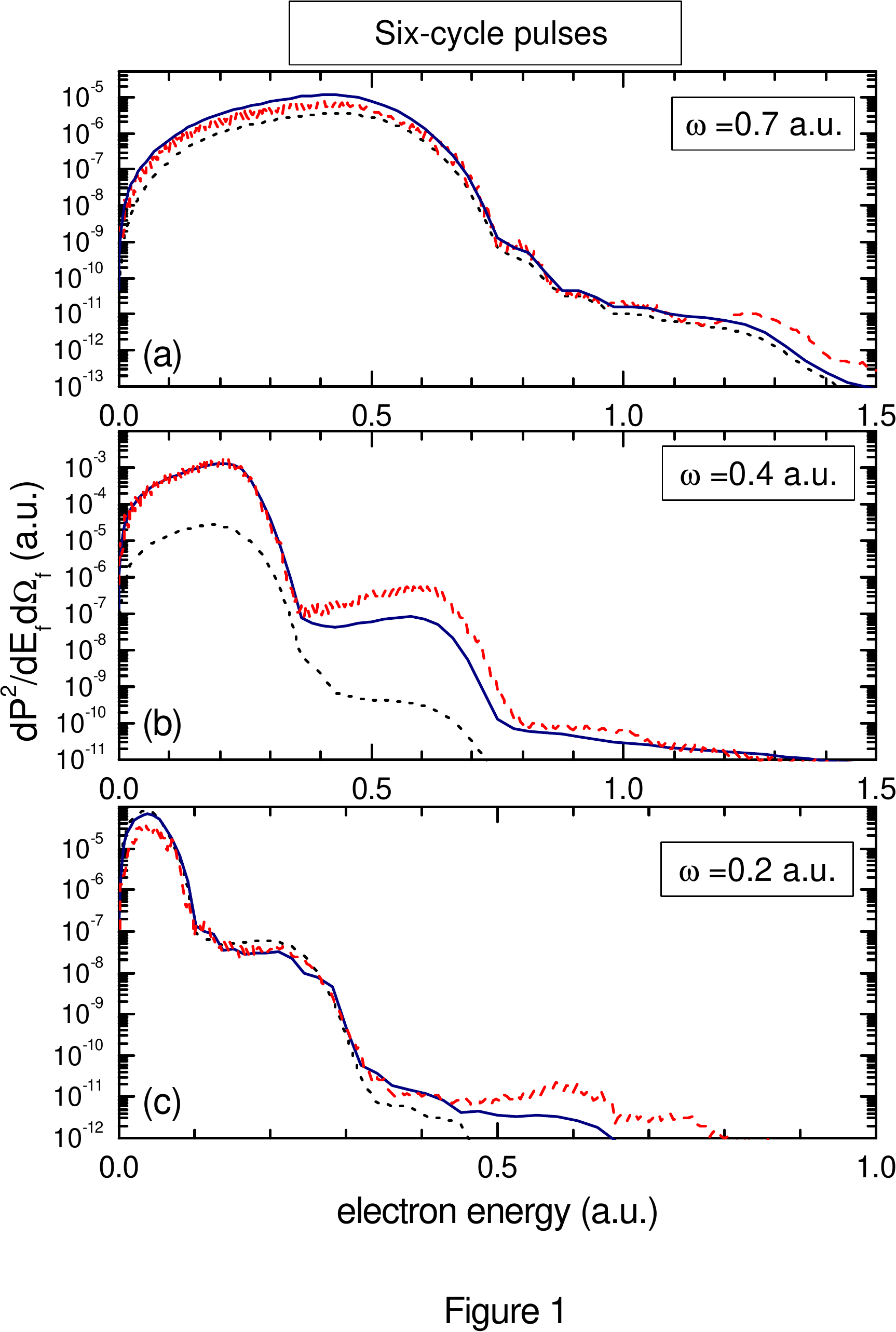}
                     \caption{\label{figure1}Differential electron emission probability, as a function
of the electron energy, for the emission angle $\theta _{e}=$ $90^{\mathrm{%
\circ }}$. The parameters of the laser field are: $F_{0}=0.001$ $\mathrm{a.u.%
}$, (a) $\omega =0.7$ $\mathrm{a.u.}$, $\tau =54$ $\mathrm{a.u.,}$ (b) $%
\omega =0.4$ $\mathrm{a.u.,}$ $\tau =95$ $\mathrm{a.u.}$and (c) $\omega =0.2$
$\mathrm{a.u.,}$ $\tau =190$ $\mathrm{a.u.}$ Solid (blue) line, SJV results;
dashed (red) line, TDSE\ values, and dotted line, results of the IJV model.}
         \end{center}
 \end{figure}

\begin{figure}[!htb]
         \begin{center}
                     \includegraphics*[width=.45\textwidth]{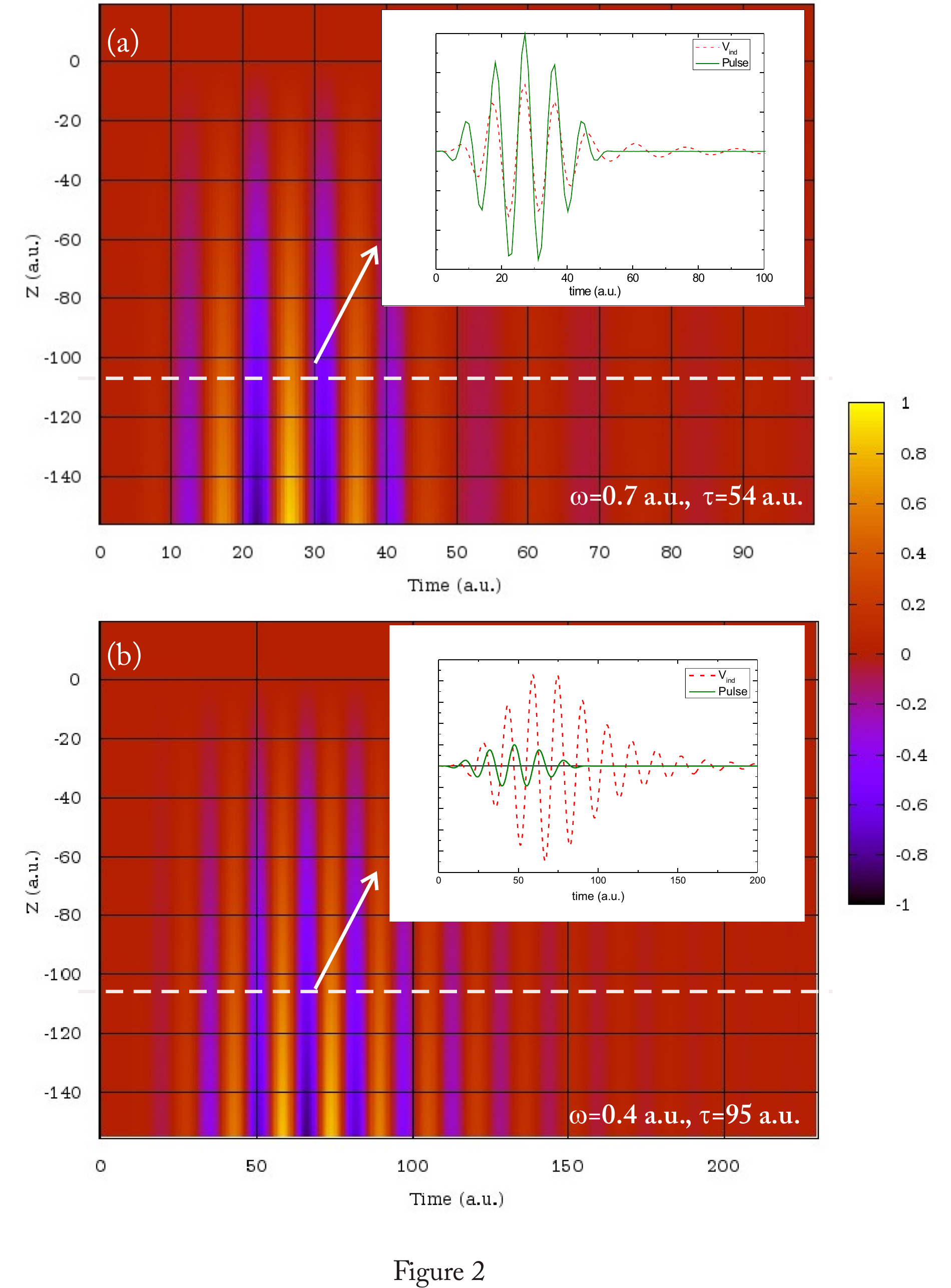}
                     \caption{\label{figure2} 2D-Representation of\ the induced potential, as a
function of time and space. Laser pulse parameters are similar to Fig.\ref{figure1}(a) and (b). Inset figures correspond to a given position inside the solid,
with solid line, the laser pulse curve, and dashed line, the induced
potential.
}
         \end{center}
 \end{figure}

In Fig.\ref{figure1}, six-cycle laser pulses with three different frequencies were
considered: $\omega =0.7$, $0.4$ and $0.2$ $\mathrm{a.u..}$ In all the
cases, to analyze the effect of the surface response on the electronic
spectra SJV and TDSE values were compared to data derived within the
previous IJV approach \cite{Fara07}, which neglects the contribution of $%
V_{ind}$. In Fig.\ref{figure1} (a) we show the emission probability corresponding to
the frequency $\omega =0.7$ $\mathrm{a.u}$., which is higher than the
surface plasmon frequency. For this frequency a good agreement between SJV
and TDSE results is found. The SJV curve runs very close to TDSE values,
showing only a small underestimation of TDSE\ results in high-velocity
range. Note that both theories present a broad maximum, which can be
associated with the above threshold ionization process. From the comparison
to values obtained within the IJV approximation, we observe that for this
high frequency the induced potential produces only a slight increment of the
probability at low electron energies, having small influence on the overall
electronic spectrum. However, when $\omega $ becomes resonant with the
surface plasmon frequency, as in Fig.\ref{figure1} (b), the induced potential
contributes greatly to increase the ionization probability in the whole
energy range. Energy distributions obtained with SJV and TDSE methods are
more than one order of magnitude higher than the one derived from the IJV
approach. In this case SJV results follows quantitatively well the behavior
of the TDSE curve, describing properly the positions of the multiphoton
maxima but underestimating TDSE probabilities around the second peak. Note
that in this case $V_{ind}$ does not represent a weak perturbation of the
laser field, as shown in Fig.\ref{figure2}(b) and it might be the origin of the
observed discrepancy. In Fig.\ref{figure1} (c) we plot the emission probability for a
laser field with a frequency $\omega =0.2$ $\mathrm{a.u.}$, lower than the
plasmon one. Again, as in Fig.\ref{figure1} (a) SJV and TDSE results run very close to
each other, displaying almost no differences with the IJV theory, which does
not contain the induced potential. This indicates that the induced potential
strongly affects emission spectra for frequencies resonant with the surface
plasmon frequency, while for small deviations from this frequency it plays a
minor role in the multiphoton ionization process.

With the aim of examinating in detail the contribution of the induced
potential, in Fig.\ref{figure2} we plot $V_{ind}$ as a function of time, for a given
position inside the solid and for the frequencies of Fig.\ref{figure1}(a) and (b). We
observe that for $\omega =0.7$ \textrm{a.u.} the induced potential tends to
follow the oscillations of the external field and its intensity steeply
diminished when the pulse is turned off. Then, in this case the collective
response of the medium produces only a weak effect on the electronic
spectrum, as shown in Fig.\ref{figure1}(a). Whereas for laser frequencies near to $%
\omega _{s}$ (Fig.\ref{figure2} (b)) the process is dominated by the induced potential,
which produces an increment of the emission probability, as observed in Fig.\ref{figure1}(b).

\begin{figure}[!htb]
         \begin{center}
                     \includegraphics*[width=.45\textwidth]{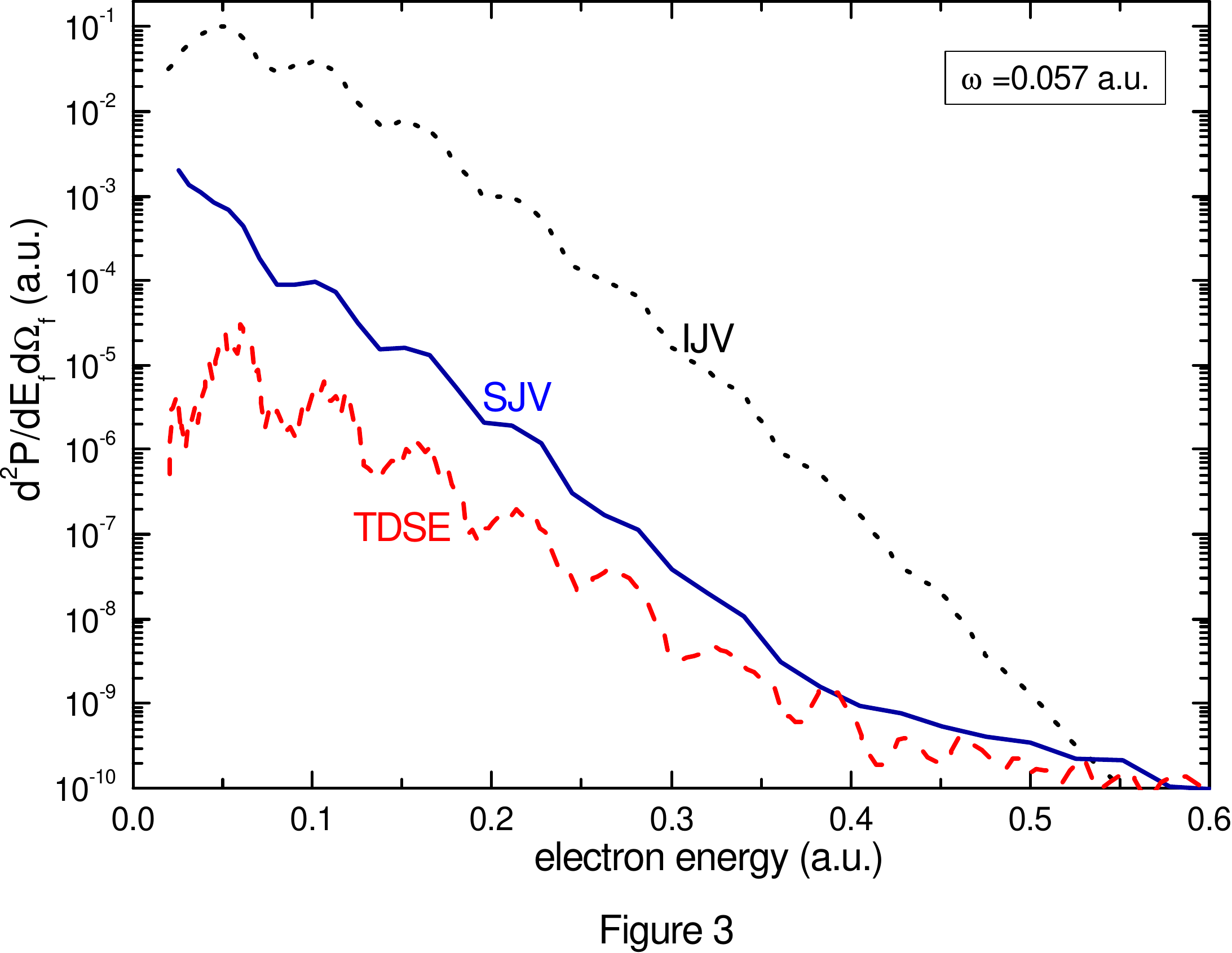}
                     \caption{\label{figure3}Similar to Fig.\ref{figure1}. Laser field with $F_{0}=0.001$ $\mathrm{%
a.u.,}$ frequency $\omega =0.057$ $\mathrm{a.u.}$ and duration $\tau =660$ $%
\mathrm{a.u.}$}
         \end{center}
 \end{figure}

Finally, in Fig.\ref{figure3} we study a six-cycle laser pulse with the frequency
corresponding to the experimental value for the Ti:sapphire laser system 
\cite{Miaja08} (\ $\omega =0.057$ \textrm{a.u.}). For this low frequency,
almost one order of magnitude lower than the plasmon one, the surface
response approximates to the static limit and electronic fluctuations screen
strongly the external field inside the solid. By comparing SJV and IJV
results it is observed that in this case the induced potential contributes
to reduce markedly the emission probability, up to two orders of magnitude
at low electron energies.\ On the other hand, it should be noted that
although the SJV theory describes properly the positions of multiphotonic
maxima, it overestimates the emission probability given by the TDSE\ method.
Such a discrepancy, which arises when $\omega $ is lower than the mean
energy of initial bound electrons, was also observed for other Volkov-type
methods applied to photoionization of atomic targets \cite{Macri03}.

\begin{figure}[!htb]
         \begin{center}
                     \includegraphics*[width=.45\textwidth]{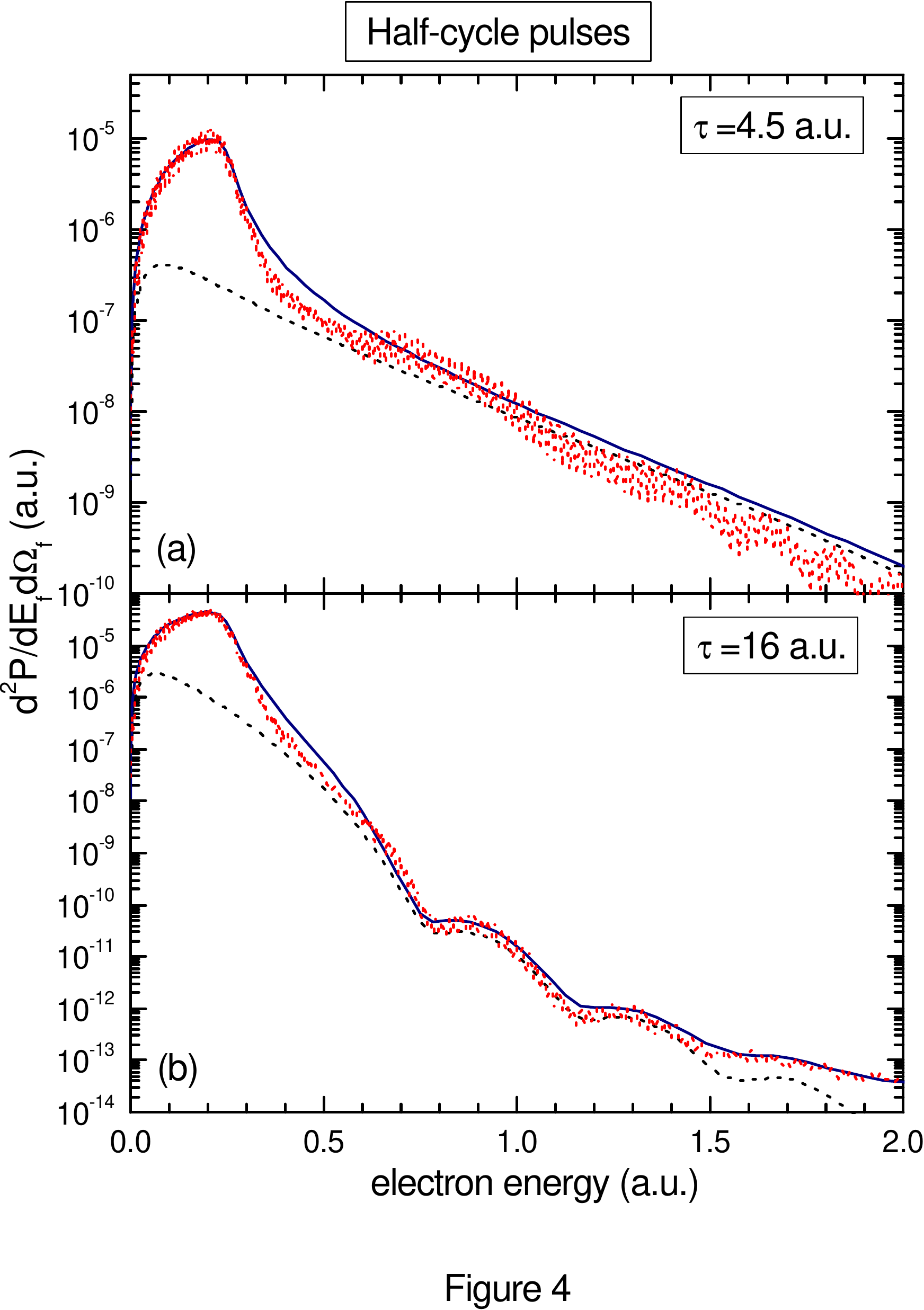}
                     \caption{\label{figure4}Similar to Fig.\ref{figure1}. Half-cycle pulse with $F_{0}=0.001$ $%
\mathrm{a.u.,}$ (a) frequency $\omega =0.7$ $\mathrm{a.u.}$ and duration $%
\tau =4.5$ $\mathrm{a.u.,}$ (b) frequency $\omega =0.2$ $\mathrm{a.u.}$ and
duration $\tau =16$ $\mathrm{a.u..}$}
         \end{center}
 \end{figure}

\begin{figure}[!htb]
         \begin{center}
                     \includegraphics*[width=.45\textwidth]{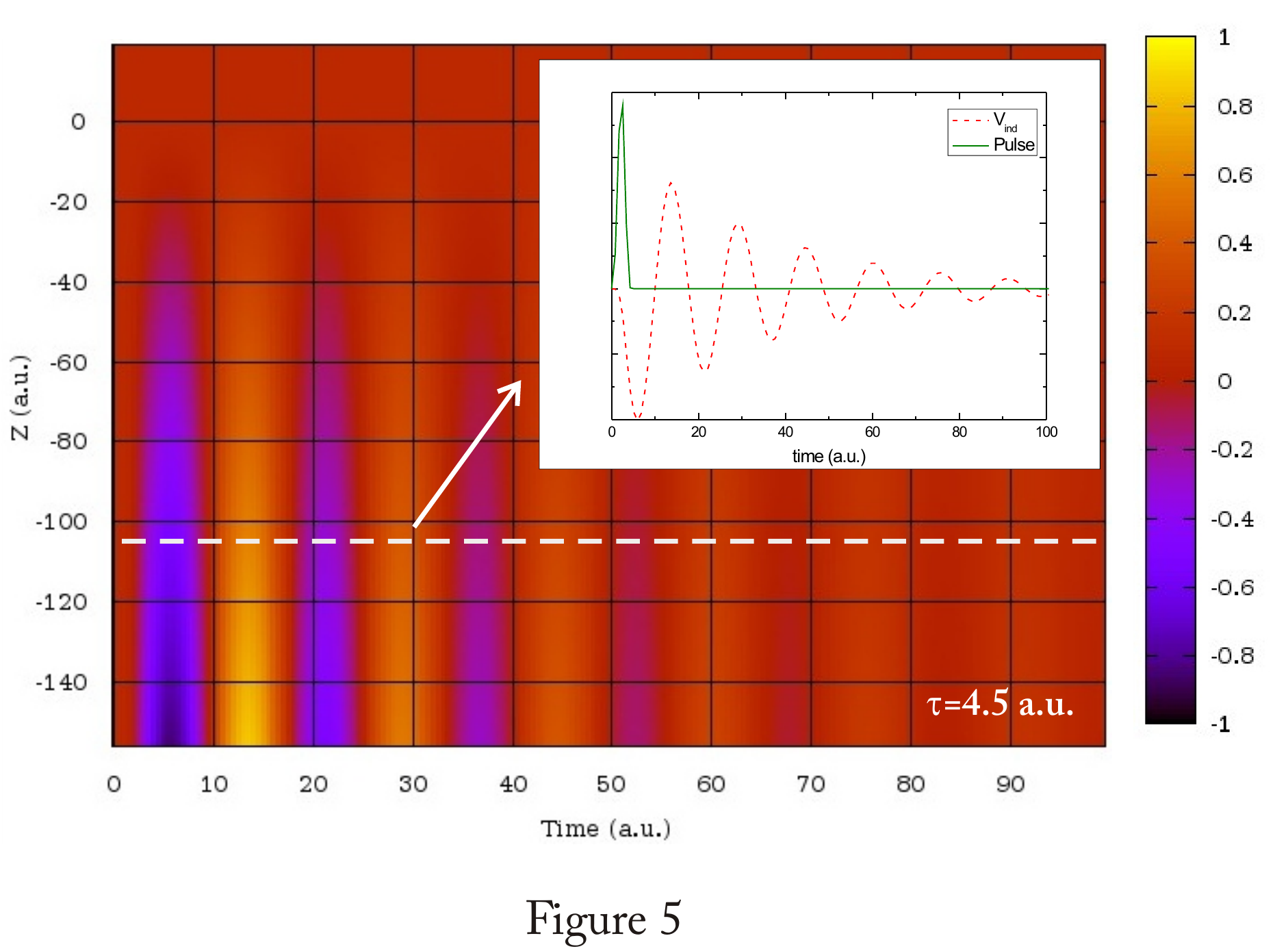}
                     \caption{\label{figure5}Similar to Fig.\ref{figure2}. Laser field with $F_{0}=0.001$ $%
\mathrm{a.u.,}$ $\omega =0.7$ $\mathrm{a.u.}$ and $\tau =4.5$ $\mathrm{a.u..}
$}
         \end{center}
 \end{figure}

To complete the previous analysis we reduce the duration of the pulse in
order to investigate the contribution of the induced potential for
photoelectron emission in the collisional regime \cite{Macri03}. In this
regime, associated \ with half-cycle pulses, the electromagnetic field does
not oscillate, producing a perturbation similar to the one resulting of the
interaction with a swift ion impinging grazingly on the surface (collision
process). \ Notice that for such ultrashort pulses the carrier frequency \ $%
\omega $ loses its meaning and the pulse\ can be characterized by the sudden
momentum transferred to the ejected electron, $\Delta p=-A^{-}(0)/c\simeq
F_{0}\tau /2$ \cite{Arbo09}. In Fig.\ref{figure4} we plot electron distributions for
half-cycle pulses with two different durations $\tau =4.5$ $\ $and $16\ 
\mathrm{a.u}$. In both cases we found a good agreement between SJV and TDSE
methods in the whole electron velocity range. Both theories present a
pronounced maximum at low electron velocities, which does not appear in the
electron distribution derived from the IJV approach, being produced by the
induced potential. To understand the origin of this increment of the
probability at low electron energies, in Fig.\ref{figure5} we plot again the induced
potential for a given position inside the solid, now for the case of Fig.\ref{figure4}
(a). \ We observe that for half-cycle pulses, without oscillations, after
the pulse has finished the induced potential still affects solid electrons
during at least a hundred atomic units more. This effect is the main source
of electrons emission at low velocities.

\subsection{Conclusions}

In the present work we have introduced the SJV approximation, which allowed
us to investigate the effects of the induced potential on the electron
emission process. The proposed theory was compared to values derived from
the numerical solution of the corresponding TDSE, displaying a good
description of the main characteristics of photoemission spectra in the
whole range of studied frequencies and durations of the laser pulse. From
the comparison between SJV probabilities and those derived from the previous
IJV approach, which does not include $V_{ind}$, we conclude that the induced
potential can play an important role in laser-induced electron emission from
metal surfaces, as expected. For laser pulses with several oscillations
inside the envelope, we found that the induced potential produces a
considerable increment of the probability when the laser frequency is
resonant with the surface plasmon one, but as $\omega $ diminishes tending
to the static case, the surface electronic density shields the laser field
inside the solid, \ leading to a markedly reduction of the photoemission
process. In addition, for electromagnetic pulses in the collisional regime,
the contribution of the surface induced potential after the pulse turns off
gives rise to a maximum in the emission spectrum at low energies.

\section{Acknowledgment}

This work was done with the financial support of Grants UBACyT, ANPCyT and
CONICET of Argentina. I. A, A. A. and V. M. S. gratefully acknowledge
financial support by GU-UPV/EHU (grant no. IT-366-07) and MCI (grant no.
FIS2007-66711-C02-02). The work of V.M.S is sponsored by the IKERBASQUE
foundation.

\end{document}